\documentclass[epsf,preprint,preprintnumber,onecolumn,12pt]{revtex4}
\usepackage{graphicx}
\usepackage[usenames,dvips]{pstcol}
\usepackage{color}
\usepackage{amssymb}
\usepackage{epsfig}
\usepackage{footnote}
\usepackage{longtable}

\newrgbcolor{maroon}{.45 0.1 0.1}
\def\pbarp{{\bar p}p}
\begin{document}
%
\preprint{EFI-04-42}

\title{ \Large A 96-Channel FPGA-based Time-to-Digital Converter}

\author{Mircea Bogdan}
\affiliation{Enrico Fermi Institute, University of Chicago}
\author{Henry Frisch}
\affiliation{Enrico Fermi Institute, University of Chicago}
\author{Mary Heintz}
\affiliation{Enrico Fermi Institute, University of Chicago}
\author{Alexander Paramonov}
\affiliation{Enrico Fermi Institute, University of Chicago}
\author{Harold Sanders}
\affiliation{Enrico Fermi Institute, University of Chicago}
%

\author{Steve Chappa}
\affiliation{Fermilab National Accelerator Laboratory}
\author{Robert DeMaat}
\affiliation{Fermilab National Accelerator Laboratory}
\author{Rod Klein}
\affiliation{Fermilab National Accelerator Laboratory}
\author{Ting Miao}
\affiliation{Fermilab National Accelerator Laboratory}
\author{Peter Wilson}
\affiliation{Fermilab National Accelerator Laboratory}
%

\author{Thomas J. Phillips}
\affiliation{Duke University}
%
%

\begin{abstract}
We describe an FPGA-based, 96-channel, time-to-digital converter (TDC)
intended for use with the Central Outer Tracker (COT)~\cite{COT} in
the CDF Experiment~\cite{CDF_TDR} at the Fermilab Tevatron. The COT
system is digitized and read out by 315 TDC cards, each serving 96
wires of the chamber. The TDC is physically configured as a 9U VME
card.  The functionality is almost entirely programmed in firmware in
two Altera Stratix FPGA's.  The special capabilities of this device
are the availability of 840 MHz LVDS inputs, multiple phase-locked
clock modules, and abundant memory.  The TDC system operates with an
input resolution of 1.2 ns, a minimum input pulse width of 4.8 ns and
a minimum separation of 4.8 ns between pulses. Each input can accept
up to 7 hits per collision.  The time-to-digital conversion is done by
first sampling each of the 96 inputs in 1.2-ns bins and filling a
circular memory; the memory addresses of logical transitions (edges)
in the input data are then translated into the time of arrival and
width of the COT pulses.  Memory pipelines with a depth of 5.5 $\mu$s
allow deadtime-less operation in the first-level trigger; the data are
multiple-buffered to diminish deadtime in the second-level
trigger. The complete process of edge-detection and filling of buffers
for readout takes 12 $\mu$s.  The TDC VME interface allows a 64-bit
Chain Block Transfer of multiple boards in a crate with transfer-rates
up to 47 Mbytes/sec. The TDC also contains a separately-programmed
data path that produces prompt trigger data every Tevatron crossing.
The trigger bits are clocked onto the P3 VME backplane connector with
a 22-ns clock for transmission to the trigger.  The full TDC design
and multi-card test results are described. The physical simplicity
ensures low-maintenance; the functionality being in firmware allows
reprogramming for other applications.
\end{abstract}
\maketitle

\section{Introduction}
The Collider Detector at Fermilab (CDF), is a large (5000-ton)
detector of particles produced in proton-antiproton collisions at 1.96
TeV at the Fermilab Tevatron~\cite{CDF_TDR}. The detector consists of
a solenoidal magnetic spectrometer surrounded by systems of segmented
calorimeters and muon chambers. Inside the solenoid, precision
tracking systems measure the trajectories of particles; the particle
momenta are measured from the curvature in the magnetic field and the
energy deposited in the calorimeters. The tracking systems consist of
a silicon-strip system with $>$750,000 channels around the beam-pipe,
followed by the Central Outer Tracker (COT), a large cylindrical drift
chamber with 30,240 sense wires arranged in 96 layers divided into 8
``superlayers" of 12 wires each~\cite{COT}. Four of the layers have
the wires parallel to the beam axis; the remaining four are tilted by
$\pm2$ degrees to provide small-angle stereo for 3D reconstruction of
tracks. The maximum drift time of the COT is ~200 ns; the maximum
drift length is 0.88 cm.

During the present Run II, which started in 2001, the peak luminosity
of the Tevatron has grown to over $10^{32}$ cm$^{-2}$sec$^{-1}$, a
factor of more than five higher than in Run I. The Tevatron operates
with a bunch spacing of 396 ns, with the result that the occupancy
(hits/channel) in the COT increases with luminosity as the average
number of proton-antiproton collisions per bunch crossing is now
substantially greater than one.  The increased occupancy will decrease
the rate at which CDF can record events to permanent media.  In
addition, although CDF is presently operating at its design readout
bandwidth, there are still more events of physics interest than can
presently be logged, particularly due to the success of the Silicon
Vertex Trigger (SVT), which has given CDF an expanded capability to
trigger on events from b-quarks.  A broad range of efforts are
underway to upgrade the readout bandwidth to allow operation at
luminosities up to $3\times10^{32}$ cm$^{-2}$sec$^{-1}$, including the
development of a new time-to-digital converter (TDC) for the COT.

\begin{figure}[!t]
\centering
\includegraphics[angle=0,width=5.5in]{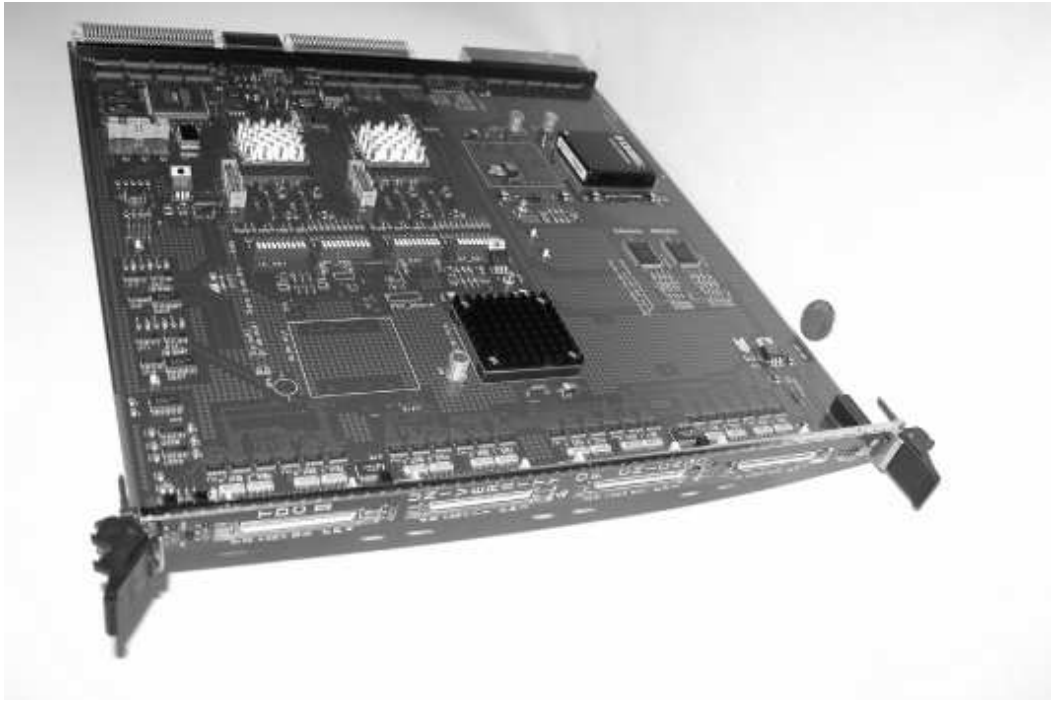}
\caption{The CDF-II TDC board. The two large chips with silver heat
sinks are the TDC FPGA's; the large black objects are DC-to-DC
converters (the layout allows addition DC-to-DC converters, not needed
and hence not stuffed on this board). The FPGA for the VME interface
can be seen in the upper left-hand corner. Connector headers and dip
switches near the center of the board facilitate debugging with a
logic analyzer, if necessary.}
\label{fig:board_picture}
\end{figure}

In this note we describe the design of a new TDC for the COT based
entirely on field-programmable gate arrays (FPGA's). Thirty working
prototypes have been built and tested. A comprehensive suite of test
routines, including some that exploit the capabilities of large FPGA's
to implement sources of test data, has been implemented and
documented. We present results on performance and readout bandwidth.

\newpage

The use of large FPGA's allows a board layout with very few chips, as
all of the data-processing functions of the TDC are contained in the
two FPGA's, each of which handles 48 channels.  A third, smaller, FPGA
serves as the VME interface. The other chips on the board are delay
lines, buffers on the input and output signals to the connectors, and
DC-to-DC converters to supply voltages not available in the existing
VME~\cite{ANSIVIPA,CDFVME} crates. The exceptionally low chip count, few
internal connections, all digital-nature, and extensive diagnostic
capabilities make the boards easy to test and
maintain. Figure~\ref{fig:board_picture} shows one of the 30
preproduction boards.

\section{TDC Specifications}
A summary of the TDC physical and operational characteristics is given
in Table ~\ref{table:specs}. The schematics of the board are available
at ~\cite{Schematics}. Details of how the TDC operates 
are given in the text below.

\begin{longtable}[c]{|l|l|l|}
\caption{ The physical and operational characteristics of the CDF-II
  TDC. \label{table:specs} }\\

\hline\hline
\bf  \small ~~~Characteristic & \bf \small ~~~Values & \bf \small ~~~Comment \\
\hline\hline
\endfirsthead
\multicolumn{3}{c}%
{ \small \tablename\ \thetable{} -- continued from previous page} \\
\hline
\bf \small ~~~Characteristic & \bf \small ~~~Values & \bf  \small ~~~Comment \\
\hline
\endhead
\hline

\multicolumn{3}{c}{ \small Continued on next page} \\ 
\endfoot

\hline \hline
\endlastfoot

\multicolumn{3}{|c|}{ TDC Digitization Performance}\\
\hline
\small Channels                & \small 96          &\small 48/FPGA \\
\small Time Bin Size           & \small 1.2 ns      & \\
\small Hits/Channel            & \small $\le$7 hits & \small Configurable via VME \\
\small Min Time Between Hits   & \small 4.8 ns      & \\               
\small Full Scale Range        & \small 304.8 ns    & \small 254 $\times$ 1.2 ns \\
\small Min Width/hit           & \small 4.8 ns      & \\
\small Max Width/Hit           & \small 304.8 ns    & \\
\small Non-Linearity           & \small $<2$ counts &  \small $<1$ count typ.~\cite{nonlinearity} \\
\small L1 Pipe-Line Size       & \small 512 words/6.144 $\mu$s & 480-bit words \\
\small Test Data RAM size      & \small 512 words/6.144 $\mu$s & 480-bit words \\
\small L2 Buffer Length        & \small $\le$64 words/768 ns    & \small 
480-bit words; Configurable  \\
\small Processing  Time        & \small 12$\mu$s after L2A    & \small
 Includes readout packing \\
\small Min interval between L2A's & \small 12 $\mu$s & L2A is Level-2 Accept\\
\hline
\multicolumn{3}{|c|}{ Prompt (Trigger) Outputs}\\
\hline
\small XFT Trigger bits/wire	& \small 6		& \small From 11 time windows \\
\small XFT Time-window  	& \small 6 ns 		& \small Min
 1.2 ns; Max 12 ns \\
\small \# of XFT Time-windows	& \small  11/wire	& \small
 Mapped into 6 trigger bits\\
\small Trigger Latency		& \small 80 ns after BC & \small First word out \\
\small Trigger Output freq.	& \small 32 bits/ 22 ns & \small See Fig.~\ref{fig:xft_output_timing} \\
\small Trigger Output		& \small 43 pins        & \small TTL, on VME P3  \\
\hline
\multicolumn{3}{|c|}{ Readout Characteristics}\\
\hline
\small VME Interface           & \small VME64  & \small Implemented in FPGA \\
\small VME Readout Modes       & \small A32/D32, A32/D64   & \small D64 in CBLT mode only \\
\small CBLT64 Transfer Rate    & \small 47 MBytes/s  & Burst speed \\
\small Test Modes              & \small  Data Generator & \small
Internal 8192 word memory\\
\hline
\multicolumn{3}{|c|}{ Physical Characteristics}\\
\hline
\small Physical Format         & \small 9U VME            & \small ANSI VIPA~\cite{ANSIVIPA} \\
\small Power Requirements (V/A)& \small +5V/15A;~  -5V/2A   & \\
\small Input Connectors        & \small 68-pin            & Mini-D Ribbon   \\
\small Input Levels            & \small LVDS              &\small  CDF uses quasi-LVDS~\cite{QuasiLVDS}\\
\small Front Panel LEDs        & \small 1 Triple LED/FPGA  & \small Configurable in firmware \\
\small Trigger Output Connector & \small VME P3  & \\
\end{longtable}

\section{Principle of Operation}

Secondary particles from antiproton-proton ($\pbarp$) collisions
traverse the Tevatron beam pipe, the silicon-strip vertex detector,
and then the COT drift chamber volume. The charged particles ionize
the gas in the drift chamber volume; the tracks are measured from the
time of arrival of the ionization on the sense wires of the
COT~\cite{COT}.  These electrical pulses, colloquially known as
`hits', are amplified and shaped by the Amplifier Shaper Discriminator
(ASDQ) cards~\cite{COT} directly on the end-plates of the COT, and
transmitted to `Repeater cards' that drive the cables to the VME
crates on the outside of the magnet yoke that contain the TDC
boards. The TDC is used to digitize the time of arrival and, as a
measure of pulse height, the width of the pulses from the Repeater
cards.

The Time-to-Digital conversion is implemented with two Altera Stratix
FPGA's, each handling 48 sense wires. This device has an LVDS
differential I/O interface that consists of one dedicated
serializer/deserializer circuit~\cite{Stratix} for each of the 48
differential I/O pairs. Serial data are received along with a
low-frequency clock. An internal phase-locked loop (PLL) multiplies
the incoming clock by a factor of 1 to 10.  Each input signal is
sampled at the resulting high-frequency clock rate, converting it
to a (1-bit) serial stream, which is then shifted serially through a
shift register. The shift register is read out as a parallel word at
the low frequency clock rate, thus converting the serial data stream
into a parallel data stream that contains the input data sampled at
the higher clock rate.  In this application the low-frequency clock is
internally generated with a 12 ns period and the multiplier factor is
set to 10, for a resulting 1.2 ns sampling rate of the incoming LVDS
signal and a 10-bit wide parallel data stream.

\begin{figure}[!ht]
\centering
\includegraphics[angle=0,width=4.5in]{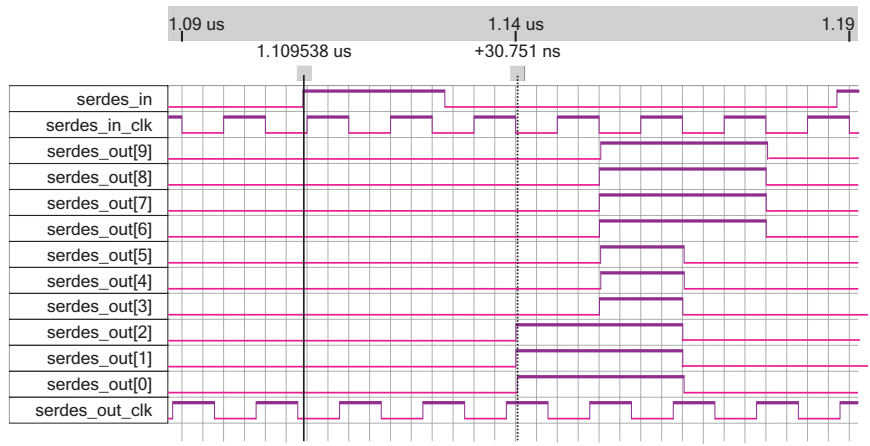}
\caption{An example of the serial-to-parallel conversion: each of the
 96 LVDS inputs is converted into a 10-bit-wide parallel data
 stream. This stream is then examined two words at a time.}
\label{fig:Principle}
\end{figure}

Figure~\ref{fig:Principle} illustrates the serial-to-parallel
conversion, as seen in the Altera Quartus II~\cite{Quartus} simulation
window. The input pulse is converted into a 10-bit parallel data
stream, (labeled as {\it serdes\_out}[9..0] in the figure) clocked at a 12 ns
period. The leading edge of a `hit' in the tracking chamber is then
determined in the Edge Detector Block by counting the number of ``0"
bits before the first ``1" bit of a string of at least four ``1" bits in
2 consecutive words at a time. The width of a hit is calculated by
counting the number of successive bits (either 0 or 1) until the start
of a string of at least four consecutive ``0" bits occurs.

\section{TDC Board - Block diagram}
\label{block_diagram}

\begin{figure}[!b]
\centering
\includegraphics[angle=0,width=4.5in]{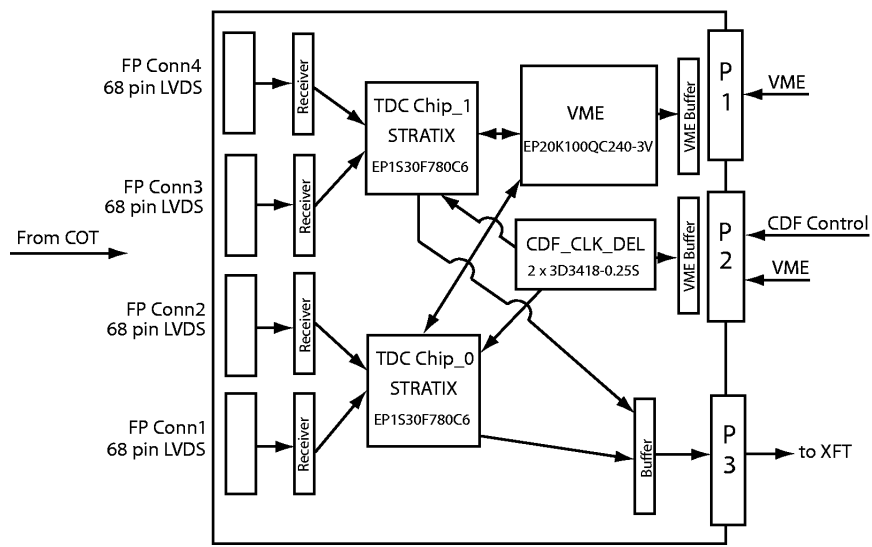}
\caption{The physical layout of the TDC board. The four input
  connectors, each with 24 LVDS channels, are on the left; the
  VME backplane connectors are on the right. The elements are
  described in turn in the text.}
\label{fig:blockdiagram}
\end{figure}

The physical layout and the data flow on the board are presented in
Figure~\ref{fig:blockdiagram}.  We step through each element in turn
below.

\begin{itemize}

\item The Front Panel (on the left in Figure~\ref{fig:blockdiagram})
  receives 96 differential inputs, arranged in four 24- channel
  connectors, which receive pulses from the amplifier/shaper circuits
  of the COT. The signals are first applied to a receiver block that
  converts them from a CDF-specific quasi-LVDS signal~\cite{QuasiLVDS}
  into standard LVDS and passes 48 of them directly to each of the two
  FPGA TDC chips, which have identical designs.

\item Each of the two TDC FPGA's (TDC Chip's) does the time-to-digital
  conversion for 48 wires to generate the Hit- Count and Hit-Data
  results. These are stored in internal VME Readout buffers
  implemented on the chip.  The TDC Chips also generate prompt data
  for the Level-1 track trigger processor (XFT).

\item The VME interface block is implemented with an Altera Apex
FPGA~\cite{Altera_Apex}.  The block coordinates VME access to the TDC Chips
for regular and Chain Block Transfers (CBLT)~\cite{ANSIVIPA} in both
32-bit and 64-bit modes. The VME chip itself is connected to only the
16 least significant bits of the VME data bus (the TDC Chips connect
to all 32 data lines).

\item 
Control signals from the CDF data acquisition and trigger systems are
brought onto the TDC board using user-defined pins of the VME P2
backplane connector.  The control signals are bussed on the backplane
of the CDF-standard 9U VME/VIPA crate~\cite{ANSIVIPA,CDFVME} from a
CDF Tracer Card~\cite{Tracer} to each of the TDC boards.  The specific
control signals used by the TDC are as follows:


\begin{itemize}

\item The CDF system clock - this is the master reference signal for
  the CDF data-acquisition system. The clock has a 132-ns period and
  is synchronous with the accelerator RF structure.  On the TDC board
  the differential PECL signal from the backplane is first converted
  to TTL, then phase-locked, buffered and applied to the TDC chips.
  The clock signal is also applied to the TDC chips after passing
  through a pair of programmable delay lines (0.25ns granularity, 64ns
  range).


\item The Bunch Crossing signal (BC), indicating that a clock corresponds to a
crossing of proton and antiproton bunches.

\item The Bunch 0 signal (B0) marks the first proton bunch which comes once per
cycle around the Tevatron ring.

\item The Level l and 2 trigger Accept/Reject signals, as well as the
Level 2 Buffer address bits.

\item The CDF\_TDC\_CALIB pulse from the VME backplane. This is converted from
PECL to ECL and can be applied to a pair of pins on each of the four
front panel connectors. The pulse is thus sent to the
amplifier-discriminator-shaper card (ASDQ)  of the COT[1] and is used for
testing and calibration. 

\end{itemize}

\item The VME P3 backplane connector is used to transmit trigger flags
generated by the TDC to the eXtremely Fast Tracker (XFT) processor to
identify tracks in the COT for the Level-1 trigger. 

\item Each FPGA is connected to a 20-pin header so that it is easy to
  use a logic analyzer for testing and diagnosis.  Signals can be
  routed to the header by programming the FPGA firmware.
\end{itemize}

\section{The TDC FPGA Chip}

\begin{figure}[!ht]
\centering
\includegraphics[angle=0,width=4.5in]{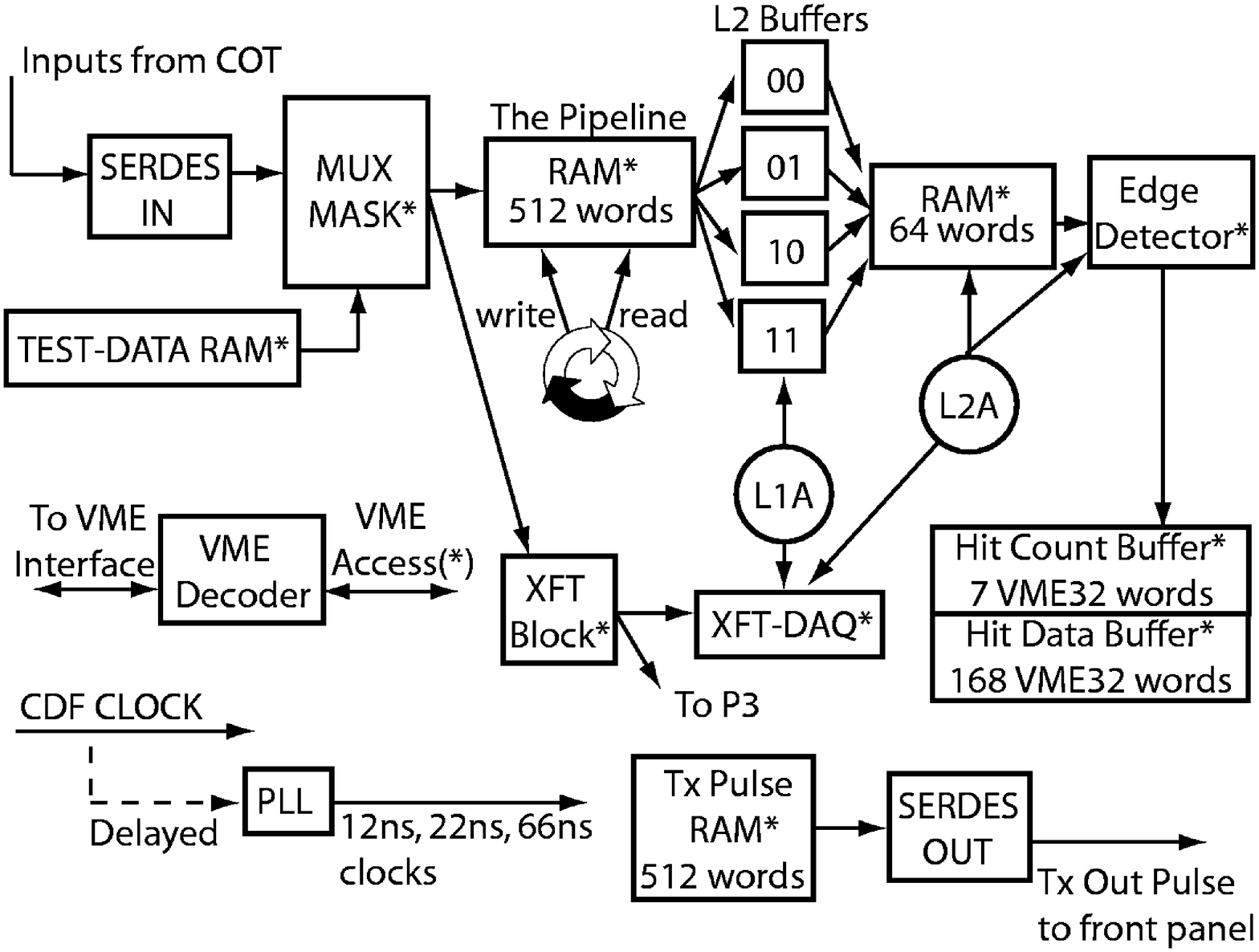}
\caption{ The functional block diagram of the TDC FPGA (`TDC Chip').
 All processing is determined by programming in firmware.  Each Chip
 handles 48 LVDS channels (shown as coming into the SERDES block in
 the upper left). The prompt trigger flags for the XFT trigger
 processor are output through the P3 VME connector.  Data are read out
 by the CDF Data Acquisition system from the Hit Count and Hit Data
 Buffers. An asterisk indicates registers or memories that are
 VME-accessible. The individual blocks in the diagram are described in
 the text.}
\label{fig:TDC_chip}
\end{figure}

The block diagram of a TDC Chip is presented in
Figure~\ref{fig:TDC_chip}.  There are two major data paths inside the
TDC Chip, one to record the COT hits for VME readout, and one for the
generation of the prompt trigger bits (`Trigger Primitives') for the
XFT trigger track processor. The Chip is also provided with a Test
Data generator, an LVDS pulse generator and a PLL clock generator.

The major functional blocks inside the TDC Chip are described
individually below. These are implemented as firmware and are
optimized for the CDF application; other applications can be
accommodated by firmware changes.

\subsection{The Input Block}
\label{muxmask}

Each TDC Chip has four banks of 12 high-speed LVDS inputs. From each
bank a 120-bit wide data bus passes data to the MUX/MASK block (see
Fig.~\ref{fig:TDC_chip}), which can be set under VME control to block
out any unwanted channel (for example, a COT wire that is continuously
set true due to a some failure). The MUX/MASK block also allows
internal testing of the TDC Chip by allowing the inputs to be switched
to a test pattern generated inside the Chip with the Test Data
Generator block (described in Section~\ref{TestDataGenerator}).  The
fast digitization and conversion to a 10-bit wide data stream for each
channel then follows.

\subsection{The Pipeline and The Level 2 Buffer System}

The CDF Level-1 trigger is deadtime-less, with all front-end data held
in a pipeline for 5.544 $\mu$s while the Level-1 trigger decision is
being made~\cite{CDFtrigger}. On the TDC card the delay is implemented
with a clocked pipeline. On receipt of a BC signal, an input memory
address counter is set to zero; the counter then increments on every
12 ns clock. The phase of this input pulse can be adjusted at each board
to compensate for signal propagation and input signal length
differences~\cite{adjust_phase}. The memory is 512 words deep, each
word containing 480 bits (48 channels $\times$ 10 bits/channel). The
memory has two ports; the first port is always writing and the second
port is always reading. The adjustable address of the second port, an
offset from the first port, is set to establish the desired delay
period for the pipeline. The maximum delay is $\sim$6.1$\mu$s.

To reduce deadtime in the CDF Level-2 (L2) trigger system, on a
Level-1 Accept signal (L1A) the data from a given beam crossing are
transferred to one of four Level-2 buffers, awaiting a L2 trigger
decision~\cite{CDFtrigger}.  The four L2 buffers are independently
controlled by an accompanying L2 buffer-selection signal. These
signals are in phase with the CDF clock pulse. The L2 buffers are
two-port memories; each has a respective write address counter for the
input write port and all share a single address counter for their
(output) read port.  This allows the writing of a second Level-1
buffer while a first is still collecting data. The write clock is
synchronous with the logic in the pipeline, 12 ns per tick. The data
are written 10-bits-wide per channel, so that 48 channels are written
at each tick in a 480-bit wide word. To achieve a maximum of 384 ns
for the time range, the maximum length is 32 words, set via VME at
initialization (the maximum drift time in the CDF COT is $\sim$200
ns).

A Level-2 Accept signal, together with an address pointer, selects one
of the four Level-2 buffers to be transferred to the Edge Detector
Block. The read port of the selected memory is driven through the full
range of stored data addresses to present all the stored data.  As not
all buffers contain data from beam crossings that pass the Level-2
trigger (and therefore these would not receive a Level-2 Accept
signal), the logic allows any Level-2 buffer to be overwritten whether
or not it has been read. No data memory buffer is ever erased.

\subsection{The Edge Detector  Block}

The purpose of the Edge Detector Block is to find hits on the wires.
 Hits are defined as pulses of at least 4.8ns in width.  Since the
pulses can be of indefinite length, the techniques of pattern matching
or look-up tables cannot be used.  The technique used in the Edge
Detector Block is to look for leading and trailing edges of a pulse.  A
leading edge is defined as a transition from low (0) to high (1) and a
trailing edge is defined as a transition from high (1) to low (0).  It
is assumed that all wires start out in a low (0) state and the first
transition to find is a leading edge.

The Edge Detector Block is made up of two modules.  The first, called
the ED, finds and stores the edges on each of the 48 wires.  The
second, called the ED48, controls the timing of the data transfer into
the Edge Detector Block, collects and packs the output hit data, and
signals when the Block is finished. Each wire has its own dedicated ED
module, making 48 on each TDC Chip; there is only one ED48 module on
each Chip.

The data from the Level-2 Buffer are fed into the Edge Detector Block
in 10- bit words.  A single ED looks for hits in two consecutive
10-bit words at a time.  The beginning of a hit is defined as a zero
followed by at least four ones, or in the case of the first word, four
ones in a row~\cite{ED_startup}.  The end of a hit is defined as a one
followed by four zeroes. There are three possible transitions in each
word and each transition needs a memory cycle.

\subsubsection{A Worked Example}

Table~\ref{tab:sampleword} shows the relationship between the position
of bits in the words sent to the ED and their respective time value as
an example in a word that has three possible transitions.

\begin{table}[!h]
\centering
\begin{tabular}{|l|r|r|r|r|r|r|r|r|r|r|}
\hline\hline
Bit Position & [9] & [8] & [7] & [6] & [5] &  [4]& [3] & [2] & [1] & [0] \\
\hline
Sample data & 0& 1& 1& 1& 1& 0& 0& 0& 0& 1\\
Time value  & 0& 1& 2& 3& 4& 5& 6& 7& 8& 9\\
\hline\hline
\end{tabular}
\caption{A sample word showing three possible transitions in the data
word sent to the Edge Detector. The first is a leading edge starting
at time value 1 and ending at time value 4, which is the second
transition.  There is also possibly another leading edge transition
starting at time value 9, depending on what is in the next word. Time
values go from early (9) to late (0).}
\label{tab:sampleword}
\end{table}

In this case, there is a leading edge (transition \#1) starting at
time value 1 and ending at time value 4 (transition \#2).  There is
also possibly another leading edge (transition \#3), starting at time
value 9, depending on what is in the next word. In this case, the ED,
which looks at 2 words at a time, would find a hit if the next
word started with three ones.

Once a hit has been found, the data describing the hit are stored and
the hit total is incremented.  Each ED has a RAM for storing the hit
data.  Each hit is characterized by the number of the time bin of the
leading edge and the width, expressed as the number of time bins in
the hit.  Thus if the data in the example of
Table~\ref{tab:sampleword} were the first word in the data stream, the
leading edge stored would be 1 and the width would be 4.  If instead
the example data were the third word, the leading edge time would be
21 (10 bits each for the first and second words plus  time value 1 in
the third word) and the width would still be 4.

\subsubsection{Maximum Number of Hits and Words}

The maximum number of words looked at per beam crossing for each wire
is variable, with a maximum of 33 words (396/12).  The maximum is set
by the time between Tevatron beam crossings, 396 ns, and the data
clocking period of 12 ns for each word (1.2 ns per bit times 10 bits
per word).  The number of words to be searched by the ED is set in a
register via VME at initialization.

Since the possibility exists that a single word could have three
transitions, the ED looks at each word three times, as shown in
Figure~\ref{fig:clock_signals}.  On a {\it clock\_0} signal, a new
data word moves into a register in the ED called {\it next\_word}.
The word that was previously in register {\it next\_word} is moved
into a register called {\it first\_word}.  The ED tries to match the
pattern "01111" with {\it first\_word}[9..5], {\it first\_word}[8..4],
{\it first\_word}[7..3] or {\it first\_word}[6..2]. These four
register segments are known as group A.  If the pattern matches, a
flag signals that the next group should look for a trailing edge
pattern "10000", otherwise the search for a leading edge continues.
On a {\it clock\_1} signal, the pattern matching is repeated, but the
register segments searched are the ones starting with bits 5-2.  These
segments are known as group B.  On a {\it clock\_2} signal, the
register segments searched are the ones starting at bits 1 and 0 and
are known as group C.  The possibility of a hit wrapping around the
end of the data word is accounted for by using the some bits in the
{\it next\_word} buffer in groups B and C.  On the next {\it clock\_0}
signal, the new data word moves into {\it next\_word}, the current
{\it next\_word} moves into {\it first\_word} and the entire process
is repeated until the last data word is moved out of {\it
first\_word}.

The maximum number of hits stored is 7.  Each ED has two small RAMs
that hold the leading edges and widths.  These RAMs are called {\it
le\_ram} (`le' for `leading edge') and {\it width\_ram}; each can hold
eight 8-bit words.  Eight-bits limits the value of leading edges and
widths to values 255 and less~\cite{explain254}.


\subsubsection{Clocks and the Edge Detector} 
\begin{figure}[!ht]
\centering
\includegraphics[angle=0,width=4.0in]{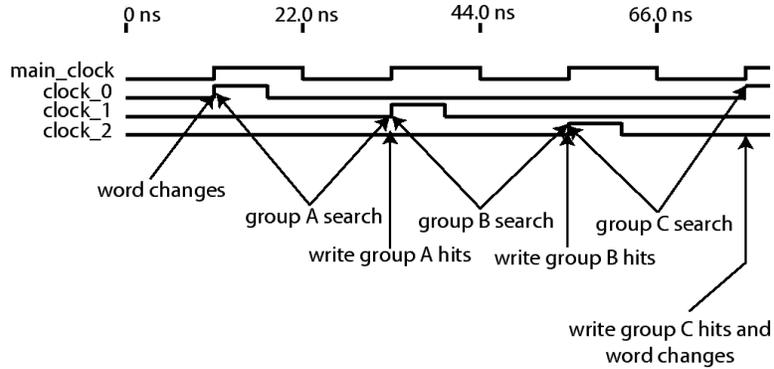}
\caption{A graphical representation of the clocks used in the Edge
Detector, and the processes that occur during each state. The word
changes are when the next 10-bit word is swapped in for edge detection.}
\label{fig:clock_signals}
\end{figure}

There are four clocks used in the ED, as described in 
Table~\ref{tab:clock_table}.
Figure~\ref{fig:clock_signals} graphically shows the clocks and the
tasks performed during the different states.

\begin{table}[!ht]
\centering
\begin{tabular}{|l|c|c|c|}
\hline\hline
Clock Name & Clock Period & Phase & Duty Cycle \\
\hline
Main Clock & 22 ns &  ~~~0 ns    & 50\% \\
Clock\_0    & 66 ns & 17.6 ns & 10\% \\
Clock\_1    & 66 ns & 39.6 ns & 10\% \\
Clock\_2    & 66 ns & 61.6 ns & 10\% \\
\hline\hline
\end{tabular}
\caption{The characteristics of the four Edge Detector clocks.}
\label{tab:clock_table}
\end{table}


\subsubsection {The Edge Detector 48 (ED48) Module}
\label{ED48}

The second part of the Edge Detector block is the ED48 module.  The
ED48 takes the data from the small (8-bit$\times$8-word) RAM in each
Edge Detector and packs the bytes into two larger RAMs.  The number of hits
found on each wire is held in the Hit Count RAM, configured in the
FPGA as 32 bits$\times$7 words.  The hit data are stored in the Hit
Data RAM, configured as 32 bits$\times$168 words (see
Table~\ref{tab:hit_data}).

\begin{table}[!ht]
\centering
\begin{tabular}{|c|c|c|c|c|}
\hline\hline
Word      & Bits 31.. 24 & Bits 23..16 & Bits 31.. 24, & Bits 23..16 \\
          & Leading edge & Width       & Leading edge  & Width\\
\hline\hline
Word 0 & \multicolumn{2}{|c|}{Hit \#1, Wire 1} &\multicolumn{2}{c|}{Hit \#2, Wire 1} \\
\hline
Word 1 & \multicolumn{2}{|c|}{Hit \#1, Wire 2} &\multicolumn{2}{c|}{Hit \#1, Wire 3} \\
\hline
... & \multicolumn{2}{|c|}{... } & \multicolumn{2}{c|}{... }\\
\hline
Word N & \multicolumn{2}{|c|}{Hit \#k, Wire j} &\multicolumn{2}{c|}{Hit 
\#(1), Wire (j+1)}\\
\hline
Word N+1 & \multicolumn{2}{|c|}{Hit \#2, Wire (j+1)} &\multicolumn{2}{c|}{Hit 
\#(3), Wire (j+1)}\\
\hline
  ... & \multicolumn{4}{|c|}{....} \\
\hline\hline
\end{tabular}
\caption{An example of the Hit Data Buffer format. Each of the 48
wires per Chip can have up to 7 hits.  Each word holds the leading
edge and widths for two hits. If a wire has no hits no entries for
that wire appear in the RAM. In the example shown, wire 1 has 2 hits,
wire 2 has one hit, and we can't tell from the table how many hits
wire 3 has.}
\label{tab:hit_data}
\end{table}

The Hit Count RAM stores the number of hits on each
wire as 4 bits, so that each 32-bit word contains the hit count for 8
wires (See Table~\ref{tab:hit_count}).  The first six words store the
information for the 48 wires per Chip.  The last word is a header word that
contains the bunch crossing counter, L2 buffer number, Chip serial
number and module ID (See Table~\ref{tab:hit_count_header}).

\begin{table}[!ht]
\centering
\begin{tabular}{|c|c|c|c|c|c|c|c|c|}
\hline\hline
Word & Bits  & Bits  & Bits  &  Bits  & Bits  & Bits & Bits & Bits  \\
     & 31-28 & 27-24 & 23-20 &  19-16 & 15-12 & 11-8 & 7-4  & 3-0 \\
\hline\hline
0 &  Wire 7 & 6 & 5 & 4 & 3 & 2 &  Wire 1 &  Wire 0\\
\hline
... & ...& ... & ... & ... & ... & ... &  ... & ...\\
\hline
5 &  Wire 47 & 46 & 45 & 44 & 43 & 42 &  Wire 41 &  Wire 40\\
\hline\hline
6 &   \multicolumn{8}{|c|}{ Header Word} \\
\hline\hline
\end{tabular}
\caption{ Hit Count Buffer format. The number of hits on each wire is
  encoded in 4 bits; there are 8 wires per word. The header word is
  described in Table~\ref{tab:hit_count_header}. }
\label{tab:hit_count}
\end{table}

\begin{table}[!ht]
\centering
\begin{tabular}{|l|l|}
\hline\hline
Bits    & Description \\
\hline\hline
$[7..0]$    & Bunch Crossing Counter \\
$[17..8]$   & $\#$ of hits in Hit Data block \\
$[19..18]$  & L2 Buffer Number \\
$[20]$      & Unused, always 0 \\
$[21]$      & Chip Serial Number: 0 for Chip 0 and 1 for Chip 1 \\
$[22]$      & TDC Type:($=1$) \\
$[31..23]$  & Module ID -set with a VME write~\cite{CDF6998}\\
\hline\hline
\end{tabular}
\caption{ Hit Count Header word. This is the last word in
Table~\ref{tab:hit_count}. The Bunch Crossing Counter, which
increments every beam crossing, is reset by the B0 signal once per
Tevatron period, and is used to verify that all data assembled into an
event come from the crossing. The Level-2 buffer number ranges from 0
to 3. The TDC Type is specific for this design, and has been
designated as Type 1. The Module ID is used in CDF to describe the
board type and serial number.}
\label{tab:hit_count_header}
\end{table}

A Level 2 Accept signal initiates the reading of data from the Level 2
Buffer by the ED48~\cite{ED_read}. The ED48 module sends a signal to each
ED to clear its buffers, and then sends a signal to the Level 2 Buffer
module to start sending data.  The signal TDC\_DONE, which is
available in a VME register, is set to {\it false} indicating that
hit-processing is in progress. At this point, the ED48 does nothing
until the 48 ED's are finished working. The ED48 module then sums the
word counts from each ED.  This sum is the number of 16-bit words to
be written in the Hit Data RAM.  Since the data are taken out of the
RAM on a 32-bit bus, the 16-bit word count is then divided by two to
get the number of 32-bit words to be read.

To move the data quickly from the ED modules to the Hit
Data RAM, the ED48 module is organized into four sections, each
holding 12 Edge Detectors.  Four 12-input multiplexers are used to
stream the data from the small ED RAMs to the Hit Data RAM in the
ED48.  This allows data to be written on each clock cycle and requires
only one extra clock cycle when changing EDs.

There are twelve 12-input and three 4-input multiplexers used to
control the flow of data from the ED modules to the RAMs in the ED48.
One 4-input and four of the 12-input multiplexers are used each for
the leading edge data, the widths, and the word counts.

A combination of counters and multiplexers control the flow of data
from the 8-word/8-bit RAMs in each ED to the Hit Data RAM in the ED48.
The number of words read from each small RAM is constant and
equal to the number of hits specified, and is independent of the
number of hits found on a wire.  If the number of hits to search for
(set by VME) is specified as six, then six words will be read from the
little RAM in each ED.  At this point, the ED48 module uses the word
count from each ED to determine when the write-enable signal on the
Hit Data RAM should be turned on and off.

The write-enable signal is controlled by the output of a compare
megafunction~\cite{Stratix} that compares the address sent to the
8-word/8-bit RAMs in the ED modules with the hit word count for the wire.

Once the ED48 is finished getting the data from each of the 48 ED
modules, it sends a signal to clear all of the ED counters and
flip-flops, and sets the signal TDC\_DONE.  This signals that the data
are ready to be read out of the memories of the ED48 module via VME.

The ED48 module controls the word counts that are written to the Hit
Data RAM with three signals, called {\it save\_wc}, {\it write\_wc},
and {\it clear\_wc}.  When the ED modules are finished, the hit count
for each wire is written to the Hit Data RAM.  The word counts are
registered onto the bus to the Hit Count RAM one clock cycle after
the EDs are done, and on the next clock cycle, are written to the RAM.
When the ED48 module is finished, this bus is also cleared.

\subsection{The XFT Block}
\label{XFT_Block}

\par The TDC XFT block generates `Trigger Primitives' used by the
eXtremely Fast Tracker (XFT)~\cite{Nils_Krumnack}, which identifies
tracks in the Central Outer Tracker (COT). The tracks are used in the
Level 1 Trigger, as well as in Level 2, and consequently the complete
pattern recognition and momentum reconstruction have to be available
within 5.5 $\mu$s after the beam crossing. The sense-wire planes of
each superlayer in the COT are tilted in the $r-\phi$ plane so that a
high momentum track will traverse each superlayer~\cite{COT}, and
consequently will travel between a pair of sense wires in each plane,
resulting in some hits that are `prompt'.  The present Run IIa XFT
splits the 396-ns interval between beam crossings into three time
bins, a {\it prompt}, a {\it not sure}, and a {\it delayed} time bin.
These three bins are logically combined to give 2 trigger bits per
wire every 396 ns. The new TDC/XFT design described here uses 11 time
windows to produce 6 output trigger bits per COT wire every 396
ns. The 6 bits, referred to below as `Trigger Primitives', are derived
from the hit occupancies in the 11 time windows with Boolean logic.
The larger number of time windows allows better momentum resolution
and fake track rejection.

The Trigger Primitives, consisting of the 6 bits per wire times 48
wires, are output each 396 ns from the XFT block on each of
the 2 FPGA's.
Every 22 ns the TDC Chip sends out 16 bits to the P3 backplane (48
bits for 66 ns, and so 48*6$=$288 bits for 396 ns).  In order to speed
up the transmission of data to the XFT, the bit for each of the (up to
6) time bins is transmitted in turn after the calculation for its
corresponding three time-windows (see below) is completed. The time
windows can be reprogrammed to optimize performance by changing data in VME
registers without firmware changes.

\par
The TDC XFT block receives two types of signals:
\begin{itemize}
\item CDF control pulses (Bunch 0 (B0), Bunch Crossing (BC));
\item The primary (digital) data-stream from the COT.
\end{itemize}
The XFT block sends out timing alignment signals and the Trigger
Primitives (trigger bits) to the P3 connector on the VME backplane for
transmission to the XFT. The TDC XFT block is controlled by
two VME registers and an internal RAM~\cite{CDF6998}. The values of the
registers determine the input and output delays, and the RAM contents
define the time-window intervals. 

\begin{figure}[!ht]
\centering
\includegraphics[angle=0,width=4.5in]{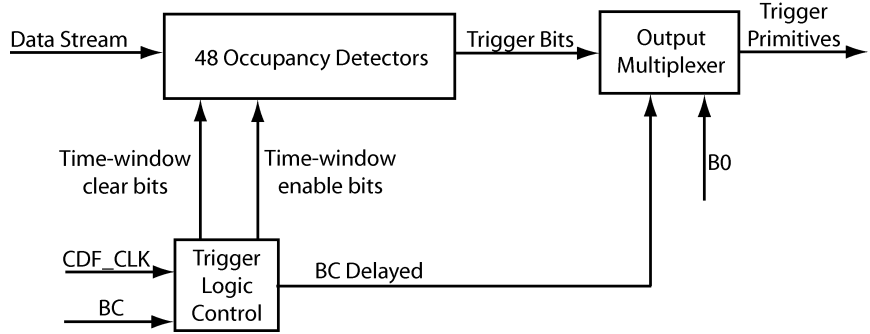}
\caption{Block diagram of the TDC XFT Logic. The data stream comes
from the MUX/MASK Block (Section~\ref{muxmask}) as 480-bit words every
12 ns. The CDF\_CLK, BC and B0 signals are the 132 ns CDF master
clock, and the Bunch Crossing and Bunch Zero signals generated by the
Tevatron, and are transmitted to the VME backplane through the CDF DAQ
system~\cite{Tracer}. The TDC XFT block sends out the Trigger
Primitives, which are the multiplexed trigger bits, and trigger
control signals.  }
\label{fig:XFT_logic}
\end{figure}

The TDC XFT block includes three main blocks (Figure~\ref{fig:XFT_logic}):
\begin{itemize}
\item Trigger Logic Control. This receives the B0 and BC pulses, and controls
the other parts of the TDC XFT;
\item 48 Occupancy Detectors. These receive the digitized COT data stream (480 
bits/12 ns), perform hit recognition, and send out 6*48 Trigger
Primitive bits for every bunch crossing to the Output Multiplexer;
\item Output Multiplexer. This receives the trigger bits from the 48
Occupancy Detectors and sends the Trigger Primitives in parallel to
the P3 backplane in 16-bit words every 22 ns.
\end{itemize}

\subsubsection{The XFT Occupancy Detector}
\begin{figure}[!ht]
\centering
\includegraphics[angle=0,width=5.5in]{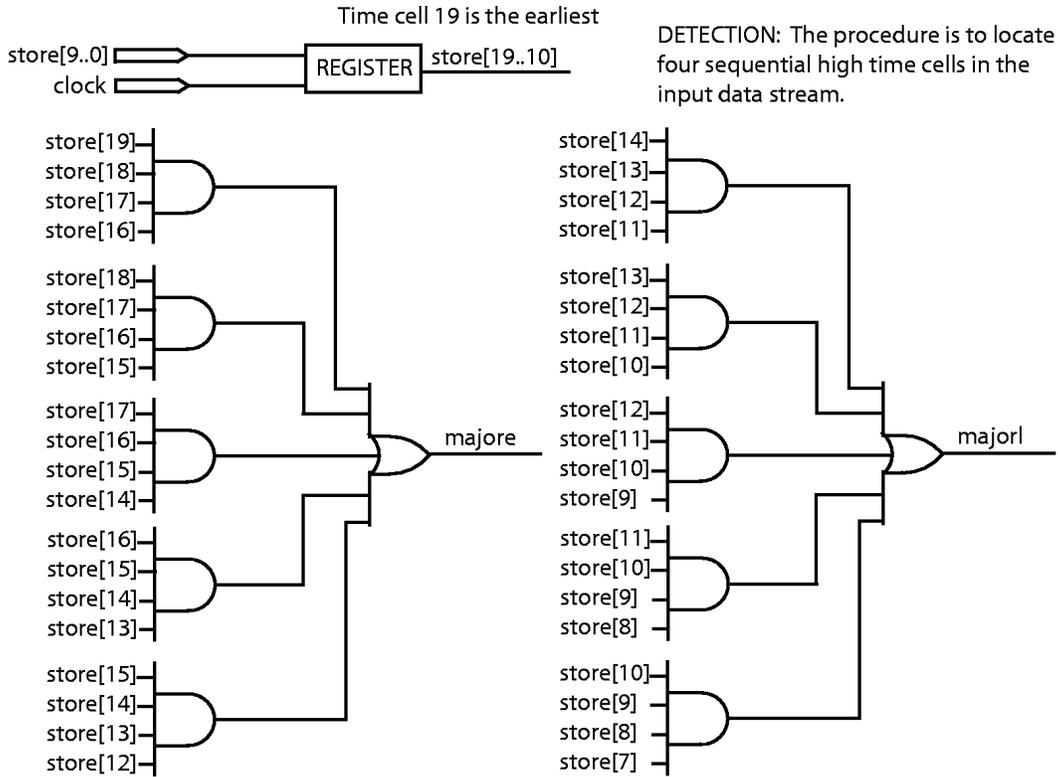}
\caption{The XFT Hit Scanner. The bus labeled ``{\it store}[9..0]" is
the input data stream for a single wire. Data come with a 12 ns
clock. The ``{\it major\_e}" (for `early') and ``{\it major\_l}" (for
`late') lines are high if a hit has occurred in the time interval
interrogated by each of the 2 Hit Scanners (the first or the last 6 ns
of every 12 ns).}
\label{fig:occupancy}
\end{figure}

Every 12 ns clock each Occupancy Detector (OD) receives a 10-bit
data-stream word.  Each OD looks for a ``1111'' pattern (a hit) in two
consecutive words in the data stream (20-bits), which corresponds to a
24 ns time interval.  The search for hits is performed separately by
two Hit Scanners inside each OD; one for hits starting in the first 5 bits
(6 ns), and the second for hits in the next 5 bits of the 20-bit data
segment. Figure~\ref{fig:occupancy} shows the block diagram of the Hit
Scanners in the Occupancy Detector.  If a hit starts in the first 6 ns
the signal labeled ``{\it major\_e}" is set high. Similarly, if a hit
starts in the last 6 ns the signal labeled {\it ``major\_l"} is set
high.  The 10-bit bus labeled ``{\it store}[9..0]" is an input data
stream for a single wire; the signal labeled ``{\it Clock}" is the 12
ns clock.


Each Occupancy Detector contains 11 independent registers (`time-window
registers') that store hit information. Each register corresponds to a
time-window. In the current firmware design each time-window is
controlled by two bits (`time-window control bits'), used to implement
separate hit scanning for the first and the last 6 ns of each 12-ns
clock interval. If the first bit is high, the register is sensitive
to hits started in the first 6 ns of every 12 ns (first 5 bits of
10-bit data word). Similarly, if the second bit is high, the register
is sensitive to hits which start in the last 6 ns of every 12 ns
(last 5 bits of 10-bit data word). As there are 33 12-ns intervals in
396 ns (the Tevatron crossing period), the time-windows for each of
the 11 registers can be defined by a 2x33 bitmap.

\begin{figure}[!t]
\centering
\includegraphics[angle=0,width=4.0in]{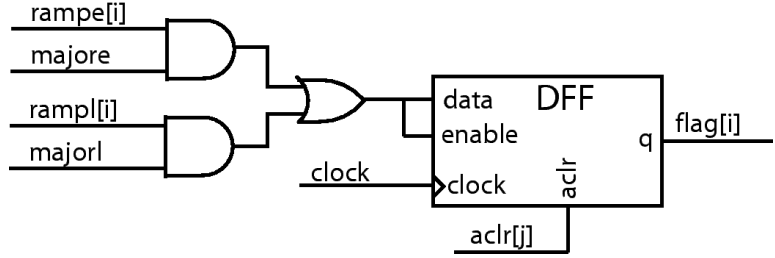}
\caption{One of the 11 Time Zone Registers. The ``{\it
  ramp\_e}[10..0]" and ``{\it ramp\_l}[10..0]" lines (``e'' for early
  and ``l'' for late) are time-window control signals (time-window
  bits) that come from the Trigger Logic Control block. The ``{\it
  major\_e}" and ``{\it major\_l}" signals correspond to the XFT Hit
  Scanner output bits. The ``{\it aclr}[6..0]" lines come from the
  Trigger Logic Control Block once per event to clear the
  corresponding group of registers. The ``{\it flag}[10..0]" bits
  correspond to an occurrence of a hit in a time region of interest,
  and become the Trigger Primitives after being processed by the
  Output Multiplexer.}
\label{fig:initial_timezone}
\end{figure}

This principle of operation allows defining time-window
ranges in units of 6 ns for the CDF XFT. However, hit scanning can be
done separately for every bit of the 10-bit data word (this would
require configuring the firmware for 10 time -window control bits per
channel). In this case the time-window unit would be 1.2
ns~\cite{11windows}. The 6 ns resolution is thought to be sufficient
for the existing XFT system. A block diagram of a time-window register
is shown in Figure~\ref{fig:initial_timezone}.

Once a register gets set to a high state, it stays high until cleared.
The Trigger Logic Control Block sends signals once per event to clear the
registers for those time windows whose Trigger Primitives have been
calculated and transmitted to the XFT.

\begin{figure}[!ht]
\centering
\includegraphics[angle=0,width=4.0in]{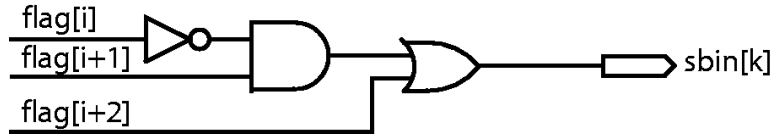}
\caption{Output Hit Logic. The ``{\it flag}[10..0]" comes from the
Initial Time Zone registers. The ``{\it sbin}[5..0]" flags are the
output trigger bits that are sent to the XFT after multiplexing. The
indices $i,i+1,i+2$ refer to individual time windows, which can be in
any temporal order. }
\label{fig:output_hit_logic}
\end{figure}

The output Trigger Primitives are computed using the data in the
time-window registers with the logic shown in
Figure~\ref{fig:output_hit_logic}. The Trigger Primitive bits
consequently depend in a (programmable) Boolean fashion on hits in the
11 time-windows. The first Trigger Primitive bit created per channel every 396
ns corresponds exactly to the result found in the first time-window
register. Subsequent bits are Boolean combinations of results
found in the later 11 windows.

%
%

\subsubsection{Trigger Logic Control}

\begin{figure}[!t]
\centering
\includegraphics[angle=0,width=4.5in]{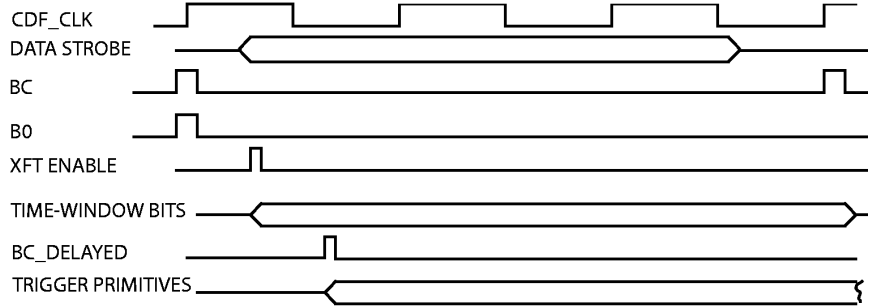}
\caption{A sample signal pattern in the Trigger Logic Control section
for a single bunch crossing. CDF\_CLK is the 132 ns clock derived from
the Tevatron. BC and B0 are the Bunch Crossing and Bunch Zero signals,
respectively. The Time-Window bits determine the logic used to make
the Trigger Primitives, which are the data sent to the XFT trigger
processor.  }
\label{fig:input_clock}
\end{figure}

\begin{figure}[!h]
\centering
\includegraphics[angle=0,width=4.5in]{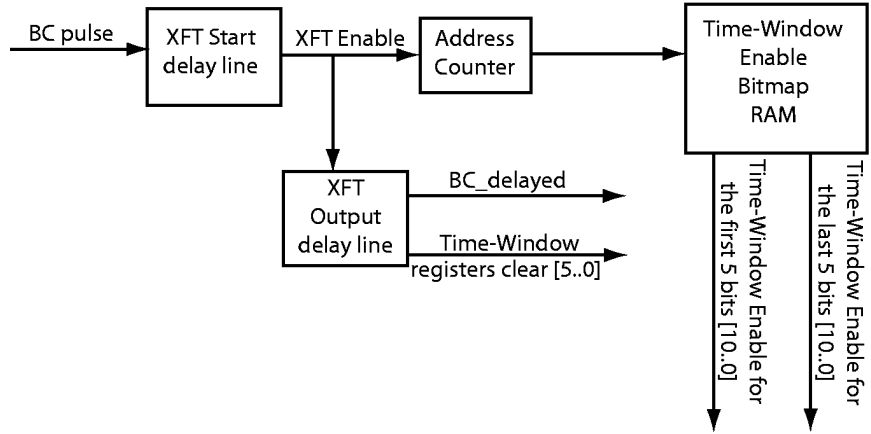}
\caption{The block diagram of the Trigger Logic Control. The ``Bunch
Crossing" (BC) line is a CDF control pulse that comes at the time the
proton and antiproton beams cross at the collision point, every 396
ns. All output pulses are synchronous with the BC pulse.  The bunches
are numbered around the 6.3-km circumference of the Tevatron ring,
with the ``Bunch\_Zero" (B0) signal being the marker of the start of the
bunch counting.  B0 is simultaneous with the BC pulse.}
\label{fig:TLC}
\end{figure}

The TDC XFT block analyzes COT data synchronously as they are
received.  The CDF control pulses Bunch Crossing (BC) and Bunch Zero
(B0) are synchronous with the data-stream.  Every 12 ns each Hit
Finder looks for hits in the input 20 bits in 6
time-windows~\cite{11windows}. Every 12 ns the Trigger Logic Control
block sets 22 bits, two per time-window, to define which 11
time-windows are current for this clock cycle. All Occupancy Detectors
get the same 22 time-window bits.  These time-window bits are stored
in a 22-bit wide RAM, which is accessible via VME. The RAM's Address
Counter is incremented every 12-ns clock and hence the time-window
bits refresh every 12 ns. The Address Counter starts to count when it
receives an XFT-Enable pulse. The XFT-Enable pulse is delayed by the
same amount as the COT data stream relative to the BC and B0 pulses. A
sample signal pattern is shown in Figure~\ref{fig:input_clock}.
The block-diagram of the Trigger Logic Control is shown in
Figure~\ref{fig:TLC}.

The ``XFT-output delay'' block sends out a ``BC\_delayed'' pulse which
enables the transmission of the Trigger Primitive bits. The
``B0\_delayed'' signal is simultaneous with ``BC\_delayed''.  The
delay of the BC pulse depends on the time-windows and can be adjusted
to send out trigger bits as soon as each trigger bit is
calculated. The delay intervals are controllable by VME-accessible
registers. When the transmission of a group of Trigger Primitive bits
(48 bits) is finished, the Trigger Logic Control Block clears the
corresponding time-window registers by sending out six time-window
clear bits (this is asynchronous).

\subsubsection{Output Multiplexer}

In each TDC Chip the Output Multiplexer (OM) receives the Trigger
Primitive bits from the Occupancy Detectors, and {\it Delayed BC} and
{\it Delayed B0} from the Trigger Logic Control Block. Each OM sends
out a 16-bit Trigger Primitive word every 22 ns on the P3 connector to
the XFT.  The signals are buffered to the P3 connector as TTL
levels. The OM also sends synchronously a Word\_0 marker (see
Table~\ref{table:multiplexer}), the B0 marker if appropriate (i.e. the
crossing is that of the Tevatron bunch 0), and an alignment signal (Data
Strobe). The OM does not perform any logical operations on the trigger
bits, but sends them in the order required by the
XFT~\cite{Nils_Krumnack}.

\begin{figure}[!ht]
\centering
\includegraphics[angle=0,width=4.5in]{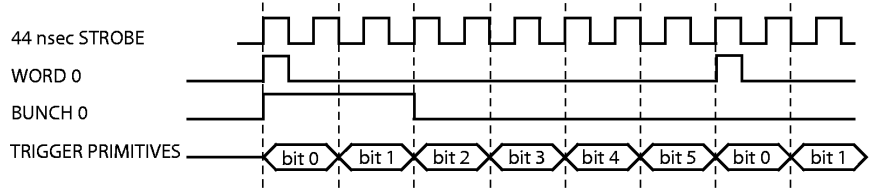}
\caption{The timing diagram of TDC signals sent to the XFT.  The 44-ns
  Data Strobe is an alignment signal, with bits being transmitted on
  both the leading and trailing edges, i.e. every 22 ns. Word Zero is
  the marker transmitted to the XFT to indicate the  start of the
  data. Bunch Zero is a signal that is present only once per Tevatron
  revolution. There are three sets of Trigger Primitive bits
  transmitted every 66 ns, as described in
  Table~\ref{table:multiplexer}.}
\label{fig:xft_output_timing}
\end{figure}

The existing cables to the XFT have too low a band-width to transmit the 22-ns
Main Clock as a data strobe, and so Trigger Primitive bits are sent on
the leading and trailing edges of a slower clock which is also
transmitted on the cables.  The Data Strobe (DS) is a 44 ns clock
formed from doubling the period of the Main Clock (see
Table~\ref{tab:clock_table}.)  Thirty-two bits, 16 from each TDC Chip,
are sent every 22 ns, as shown in Table~\ref{table:multiplexer}, with
18 such cycles in the beam-crossing period of 396 ns.
Figure~\ref{fig:xft_output_timing} shows the XFT output timing.

\begin{table}[t]
\centering
\begin{tabular}{|c|c|c|c|c|c|c|}
\hline\hline
Clock   & Pins 1-17     & Pin 18  &  Pin 19 & Pin 23         &       Pins 25-33 &    Pins 37-43\\
Cycle   & Wires, Bit      & {\footnotesize{Word0}}  &   {\footnotesize{
    B0}}   & {\footnotesize{Data\_Strobe}} &  Wires, Bit    &  Wires, Bit   \\
\hline\hline
0       &  00-15, bit 0 & High & B0    & High    &  48-56, bit 0 &  57-63, bit 0  \\
\hline
1       &  16-31, bit 0 & Low~  & B0    & Low~     &  64-72, bit 0 &  73-79, bit 0  \\
\hline
2       &  32-47, bit 0 & Low~  & B0    & High    &  80-88, bit 0 &  89-95, bit 0  \\
\hline
3       &  00-15, bit 1 & Low~  & B0    & Low~     &  48-56, bit 1 &  57-63, bit 1  \\
\hline
4       &  16-31, bit 1 & Low~  & B0    & High    &  64-72, bit 1 &  73-79, bit 1  \\
\hline
5       &  32-47, bit 1 & Low~  & B0    & Low~     &  80-88, bit 1 &  89-95, bit 1  \\
\hline
6       &  00-15, bit 2 & High  & B0    & High    &  48-56, bit 2 &  57-63, bit 2  \\
\hline
7       &  16-31, bit 2 & Low~  & B0    & Low~     &  64-72, bit 2 &  73-79, bit 2  \\
\hline
8       &  32-47, bit 2 & Low~  & B0    & High    &  80-88, bit 2 &  89-95, bit 2  \\
\hline
9       &  00-15, bit 3 & Low~  & B0    & Low~     &  48-56, bit 3 &  57-63, bit 3  \\
\hline
10      &  16-31, bit 3 & Low~  & B0    & High    &  64-72, bit 3 &  73-79, bit 3  \\
\hline
11      &  32-47, bit 3 & Low~  & B0    & Low~     &  80-88, bit 3 &  89-95, bit 3  \\
\hline
12      &  00-15, bit 4 & High  & B0    & High    &  48-56, bit 4 &  57-63, bit 4  \\
\hline
13      &  16-31, bit 4 & Low~  & B0    & Low~     &  64-72, bit 4 &  73-79, bit 4  \\
\hline
14      &  32-47, bit 4 & Low~  & B0    & High    &  80-88, bit 4 &  89-95, bit 4  \\
\hline
15      &  00-15, bit 5 & Low~  & B0    & Low~     &  48-56, bit 5 &  57-63, bit 5  \\
\hline
16      &  16-31, bit 5 & Low~  & B0    & High    &  64-72, bit 5 &  73-79, bit 5  \\
\hline
17      &  32-47, bit 5 & Low~  & B0    & Low~     &  80-88, bit 5 &  89-95, bit 5  \\
\hline\hline
\end{tabular}
\caption{ The output of the XFT Multiplexer in 6-trigger-bit mode on
  the pins of the P3 backplane connector. The table represents
  successive 22 ns time-slices as rows. The wire numbers correspond to
  the COT wire numbering scheme~{\protect{\cite{Nils_Krumnack}}}. The
  bits 0-5 are the Trigger Primitives (see Section~\ref{XFT_Block}).}
\label{table:multiplexer}
\end{table}

\subsubsection{TDC XFT-DAQ Block}

The TDC XFT-DAQ block, used for testing and diagnostic purposes only,
connects to a dedicated DAQ system similar to the hit-data stream. It
has the same structure, consisting of a `Pipeline', L2 buffers, and
VME-readout buffers, and follows the same L1A/L2A sequence as that of
the main hit-data stream. The length of the XFT-L2 buffers is also VME
controlled.  For testing, the XFT block is fitted with another simple
VME Readout RAM that contains the current XFT Trigger Primitives. This
RAM can be frozen and read out via VME for diagnostic tests, in
particular for debugging the TDC-XFT connection.

\subsection{ The Test Data Generator}
\label{TestDataGenerator}

The Test Data Pattern Generator inside the TDC Chip allows testing of
the functionality of the TDC by running test data through the entire
sequence. The Pattern Generator is implemented as a VME accessible
dual-port, 512-bit wide RAM, which can store 8192 32-bit words. The
first 480 bits of the 512 are used to drive the 48 channels with
10-bit words of test data. In Test- Data Mode, the reading of the RAM
is synchronized with the CDF\_B0 back- plane signal. The Pattern
Generator is used in the tests described below in
Section~\ref{test_results}.

\subsection{The ASDQ Pulse Generator}
\label{LVDS_Pulse_Generator}


The COT electronics chain of ASDQ-Repeater-TDC boards can currently be
calibrated by having each TDC send a differential-ECL calibration
pulse `upstream' on a pair of dedicated lines in the multi-conductor
signal cable to its associated ASDQ cards.  Each ASDQ uses this input
to generate a calibration pulse at its input, which then propagates
back through the data lines of the same signal cable~\cite{COT}.  This
loop allows a time calibration of the signal path, including the
actual cable length.  Multiple chains can be calibrated simultaneously
by providing the bussed backplane signal CDF\_TDC\_CALIB to each TDC
(see Section~\ref{block_diagram}).  The new TDC design provides for
the same functionality, but also provides for local generation on the
TDC of the pulses to the ASDQ cards.  The choice of the two
calibration modes is made by VME register selection.

In the local generation mode, the SERDES OUT block (see
Figure~\ref{fig:TDC_chip}) generates a serial LVDS pulse pattern,
available as an ECL signal on the front panel.  The timing and number
of pulses are controlled via VME by writing the contents of the Tx
Pulse Memory, which is implemented as a 10-bit wide, 512-word RAM.


\subsection{The Clock Generator Block}

The Clock Generator block, implemented in each TDC FPGA with PLL's,
generates the 12-ns and 22-ns clocks used inside the Chip. All the
clocks are synchronous with the delayed CDF clock. The prompt CDF
clock is also received and is used to latch the CDF-specific
back-plane control signals (described in Section~\ref{block_diagram}).

The four 12-ns clocks are generated onto output pins of the FPGA as
LVDS signals and routed back into the Chip onto the FPGA's dedicated
high-speed clock input pins, one for each high-speed I/O bank (see
Section~\ref{muxmask}).

\section{The VME interface block}

The VME Interface is also implemented with an Altera
FPGA~\cite{Altera_Apex}.  The design permits Chained Block Transfer
(CBLT) read commands in both 32 and 64-bit modes for data transfer.
CBLT uses geographical addresses different from the ones normally used
in the crate, recognized only by the participating modules. In CDF
there are typically 18 TDC cards per VME crate.

There are two possible CBLT read commands:

\begin{enumerate}
\setlength{\itemsep}{-0.01in}

\item Read Block Transfer from virtual slot 30: Hit Count words are
read from every TDC module in the crate.  Each TDC produces 14
words/board in 32-bit mode and 8 words/board in 64-bit mode.

\item Read Block Transfer from virtual slot 31: Hit Data words are
read from every TDC module in the crate.  Each TDC produces up to 336
words/board in 32-bit mode and up to 168 words/board in 64-bit mode.
\end{enumerate}

In this firmware implementation, the CBLT mode is enabled by default.
Up to 18 TDC cards sit in a VME crate on the CDF detector; the TDC
module closest to the Crate CPU is automatically considered first in
the chain.  The setting of a module as last in the chain, or 
the possible removal of a module from the readout, is done by writing
to a register in the module's VME Chip to disable CBLT
mode~\cite{CDF6998}.

\section{Power Block}

The TDC board receives +5V/15A and -5V/2A on the P0, P1, and P2
backplane connectors. The board generates +1.5V/15A and +3.3V/10A with
DC/DC converters, and +2.5V/3A using a linear regulator. A -3.3V
voltage is also generated and passed through the front panel
connectors to the amplifier-discriminator-shaper card
(ASDQ)~\cite{COT}. Spare locations on the TDC board are provided for
two additional DC/DC converters.

\section{Testing and Results}
\label{test_results}

The basic idea of the tests is simple: we compare the output of the
TDC boards with a reference set, which is computed automatically by
the test routines from the input data stored in the RAM in the
TDC Chips.

\par The functionality of the TDC board was checked with a suite of
test routines that exercise the Edge Detector (ED48 test), the TDC XFT
block (XFT test) and the Chain Block Transfer in both 32-bit and
64-bit modes (CBLT32/64 tests).  The following tools are required to
test the TDC boards:

\begin{itemize}
\setlength{\itemsep}{-0.01in}
\item A VME/VIPA crate with a CDF-specific P3 backplane\cite{ANSIVIPA};
\item A commercial VME crate controller\cite{MVME}, with
 VxWorks~\cite{VxWorks} and FISION~\cite{FISION} software; 
\item A CDF TESTCLK V7 board (`Testclk') ~\cite{TESTCLK};
\item A Personal Computer with FISION~\cite{FISION} software; 
\item Several TDC boards (at least two boards are needed for CBLT tests).
\end{itemize}

 All test routines are implemented as C-code, which is executed on the
PowerPC (MVME2301 or MVME5500) crate controller after compilation.
Other crate controllers can also be used without any significant
differences.

\par The results of these tests are described below in more
detail. The burst-mode readout speed achieved in CBLT64 mode is 47
MB/sec. The Edge Detector (ED48) processing time is
always less than 12 $\mu$s (for 7 hits/wire).

\subsection{The ED48 Test}
\begin{figure}[!ht]
\centering
\includegraphics[angle=0,width=2.3in]{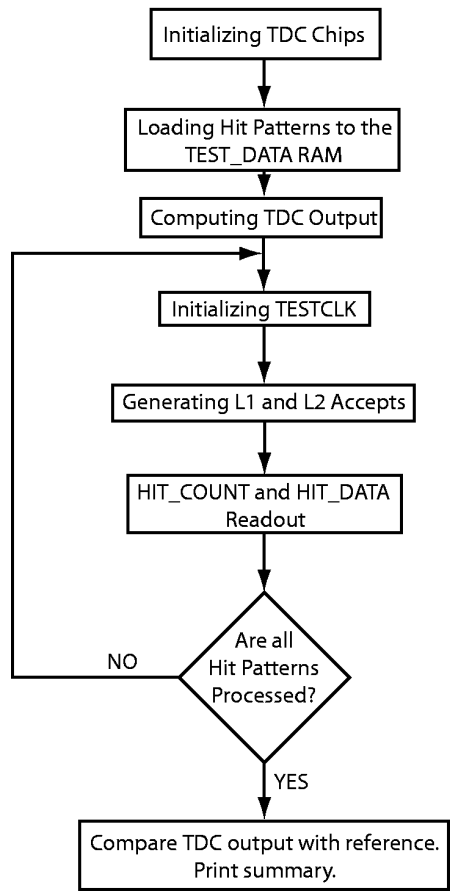}
\caption{Block diagram of the ED48 test.} 
\label{fig:ED48_test}
\end{figure}

 The ED48 test checks  the ED48 logic, internal data paths
(RAMS and buffers) and timing of the L1A, L2A, B0, and BC pulses.
A block-schema of the test is shown in Figure~\ref{fig:ED48_test}.

In the beginning of the ED48 test the values in the VME registers are
adjusted for the current L1\_accept delay (relative to B0), and the
TDC is switched to use the Test Data RAM content as an input
data-stream. A sample hit pattern is then downloaded to the RAM, which
is big enough to hold 15 events.  The test program uses the RAM
contents to compute the reference output so that it should correspond
bit-by-bit to the hardware TDC output. The Testclk Board in the VME
crate is initialized to send CDF control pulses to the backplane.  One
test consists of a set of 15 repetitions (one for every event in the
Test Data RAM). At each repetition the L1 Accept and
L2 Accept trigger control signals are generated by the Testclk with an
incremental delay of 396 ns relative to B0, so that the next event in
the Test Data RAM is processed by the ED48. At the end of each
repetition (i.e. when the {\it Tdc\_Done} bit is set- see
Section~\ref{ED48}), the Hit Data and Hit Count RAMS are
read out and the contents stored in a buffer in the Crate Master
RAM. After 15 repetitions, when all data from the  Test Data RAM
have been processed by the ED48 module, the TDC output is compared
bit-by-bit with the reference TDC output. The test is passed only if
the TDC output is in bit-by-bit agreement with the reference
data. During TDC testing the test was repeated thousands times using
different hit patterns in the Test Data RAM with no errors.

\subsection{The TDC XFT Block Test}

The TDC XFT block test is very similar to the ED48 Test.  The
difference is that the output of the TDC XFT block, the XFT Daq
RAM, is read out after every L2 Accept instead of the Hit Data
and Hit Count RAMs, and the contents compared bit-by-bit with
expectations.  The test is passed only if the outputs are in a full
bit-by-bit agreement.  The ED48 and the TDC XFT tests were executed
simultaneously using a single routine.

\subsection{Testing Chain Block Transfer: the CBLT32/64 Test}

The CBLT 32/64 tests read out multiple boards in a crate
sequentially. Tests were performed with up to 18 TDC boards in a VME
crate. The tests are designed to check the VME chain-block transfer
capabilities of the board, and also crate-wide characteristics such
the crate backplane capability, stability, and multi-board
performance. 

In the CBLT tests the initial contents of the Level 2 buffer RAMs are
used to predict the output of the ED48 modules as read from the Hit
Count and the Hit Data RAMs.  In the full-crate test 18 TDC boards are
used. For each of the 36 TDC Chips in the test (2/board), the L2
buffer lengths are set so that all possible patterns are sampled. The
order of accessing the four L2 buffers is also selected differently
for each TDC board, so that all combinations of L2 buffer and buffer
length are sampled.

Before starting the test cycles the Testclk board is initialized to generate
the CDF Clock (132 ns period), required for the PLLs on the TDCs.
  
Each test cycle first reads the contents of the Hit Count RAMs,
and then, using these word counts, the Hit Data RAMs. Both the
hit counts and the hit data are compared bit-by-bit with the predicted
reference output.  The test is then repeated for another combination
of L2 buffer sequence assignments and L2 buffer lengths.  If an error is
detected, a dump of all TDC memories is printed out. The CBLT32 and CBLT64
tests were repeated $5\times10^9$ times each without a single failure.



\section{Conclusions and Summary}

A new 96-channel TDC has been designed for the CDF Experiment at
Fermilab using the multichannel bit-sampling capabilities of the
AlteraStratix FPGA family. The board, built in a 9U VME format,
contains few other components other than the 2 TDC FPGA's, a VME
controller implemented in a 3rd FPGA, DC-to-DC-converters, and
input/output buffers. The functionality is exceptionally flexible,
being controlled by firmware, so that it can be reprogrammed for
different applications. The TDC has extensive test capabilities,
implemented directly in the FPGA's.  Thirty boards have been built and
tested. The reliability of the board is high as the chip count is very
low.  A full crate of the CDF-II TDC has been operated and read out in
64-bit block-transfer mode at a speed of 47 Mbytes/sec. 

\section{Acknowledgments}
We thank Bill Badgett, Frank Chlebana, Pat Lukens, Aseet Mukherjee,
and Kevin Pitts for help, support, and advice. Nils Krumnack and Ed
Rogers deserve special thanks for providing critical input on the XFT
specifications and the design of the XFT sections of the TDC. We thank
Rich Northrop for the picture of the board.
 
This work was supported in part by the National Science Foundation
under grant number 5-43270, and the U.S. Department of Energy.


\begin{thebibliography}{99}
\bibitem{nonlinearity} Any deviation from a linear correspondence with
 unit slope is due to the quantized nature of the sampling in time
 bins; the phase-locked loops are much more precise than 1 bin over
 the full-scale range.
\bibitem{adjust_phase} The relative delay between the clock and BC
  is set in 250 ps steps, determined by values written to a VME
  register, described in Ref.~\cite{CDF6998}.
\bibitem{ED_startup} At initialization the Edge Detector is
  programmed to assume that four zeros preceded the start of the data
  stream, so that if the first four bits examined are ``True'' a hit is
  registered at the start of the stream.
\bibitem{explain254} The largest leading edge value is actually 254,
and not 255, due to the quantization in the pattern recognition
algorithm.
A leading edge with value 255 would have the pattern "01111" in the
25th word, starting at bit 5.  This register segment is in group B.
If this leading edge were written, the possibility also exists that a
leading edge starting at bits 4-2 could also be written.  These
leading edges would have values 256, 257 and 258 respectively.  Since
those values could not fit into eight-bits, it was decided to make the
value of the largest leading edge 254.
\bibitem{ED_read} If the ED48 Module is already reading from the Level
2 Buffer, the Level 2 Accept is ignored. The $\sim$ 12 $\mu$s readout
time of the FPGA-based design described here is much shorter than the
$\sim$ 500 $\mu$s of the DSP-based design of the older TDC's currently
used by CDF. The L2 Accept rate is limited to less than $\sim$ 300 Hz,
so the deadtime incurred at 12 $\mu$s is not a problem.
\bibitem{11windows} Note that each time-window bit corresponds to a
separate Hit Scanner, which would also have to be implemented in the
firmware. This should not be hard, if ever needed.  The Occupancy
Detector has 11 time-window registers, one for every time-window.
\end{thebibliography}
\end{document}